\documentclass[prl,superscriptaddress,twocolumn,showpacs,amsmath,amssymb]{revtex4}

\usepackage{epsfig}
\usepackage{dcolumn}
\usepackage{bm}
\usepackage{ulem}

\usepackage{graphicx}

\def\simge{\mathrel{%
   \rlap{\raise 0.511ex \hbox{$>$}}{\lower 0.511ex \hbox{$\sim$}}}}
\def\simle{\mathrel{
   \rlap{\raise 0.511ex \hbox{$<$}}{\lower 0.511ex \hbox{$\sim$}}}}

\newcommand{\pt}{p_T}
\newcommand{\Et}{E_T}

\newcommand {\GeVc} {{\rm GeV}/c}

\newcommand {\GeV} {{\rm GeV}/c^2}

\newcommand{\MET}{\mbox{$E\kern-0.60em\raise0.10ex\hbox{/}_{T}$}}
\newcommand{\vecMET}{\mbox{$\vec{E}\kern-0.60em\raise0.10ex\hbox{/}_{T}$}}
\newcommand{\METrel}{\mbox{$E\kern-0.60em\raise0.10ex\hbox{/}_{T,rel}$}}
\newcommand{\MetDeltaPhi}{\Delta\phi_{E\kern-0.50em\raise0.08ex\hbox{\scriptsize /}_{T},(\ell{\rm,jet)}}}

\newcommand{\W}{W}
\newcommand{\Z}{Z}
\newcommand{\WW}{WW}
\newcommand{\WZ}{WZ}
\newcommand{\ZZ}{ZZ}

\newcommand{\HWW}{H\rightarrow WW^{(\ast)}}

\newcommand{\ppbar}{p \overline{p}}
\newcommand{\qqbar}{q \overline{q}}

\newcommand{\ttbar}{t \overline{t}}

\newcommand{\mh}{m_H}

\begin{document}

\title{ \boldmath Search for a Higgs Boson Decaying to Two $W$ Bosons at CDF}

\affiliation{Institute of Physics, Academia Sinica, Taipei, Taiwan 11529, Republic of China} 
\affiliation{Argonne National Laboratory, Argonne, Illinois 60439} 
\affiliation{University of Athens, 157 71 Athens, Greece} 
\affiliation{Institut de Fisica d'Altes Energies, Universitat Autonoma de Barcelona, E-08193, Bellaterra (Barcelona), Spain} 
\affiliation{Baylor University, Waco, Texas  76798} 
\affiliation{Istituto Nazionale di Fisica Nucleare Bologna, $^w$University of Bologna, I-40127 Bologna, Italy} 
\affiliation{Brandeis University, Waltham, Massachusetts 02254} 
\affiliation{University of California, Davis, Davis, California  95616} 
\affiliation{University of California, Los Angeles, Los Angeles, California  90024} 
\affiliation{University of California, San Diego, La Jolla, California  92093} 
\affiliation{University of California, Santa Barbara, Santa Barbara, California 93106} 
\affiliation{Instituto de Fisica de Cantabria, CSIC-University of Cantabria, 39005 Santander, Spain} 
\affiliation{Carnegie Mellon University, Pittsburgh, PA  15213} 
\affiliation{Enrico Fermi Institute, University of Chicago, Chicago, Illinois 60637} 
\affiliation{Comenius University, 842 48 Bratislava, Slovakia; Institute of Experimental Physics, 040 01 Kosice, Slovakia} 
\affiliation{Joint Institute for Nuclear Research, RU-141980 Dubna, Russia} 
\affiliation{Duke University, Durham, North Carolina  27708} 
\affiliation{Fermi National Accelerator Laboratory, Batavia, Illinois 60510} 
\affiliation{University of Florida, Gainesville, Florida  32611} 
\affiliation{Laboratori Nazionali di Frascati, Istituto Nazionale di Fisica Nucleare, I-00044 Frascati, Italy} 
\affiliation{University of Geneva, CH-1211 Geneva 4, Switzerland} 
\affiliation{Glasgow University, Glasgow G12 8QQ, United Kingdom} 
\affiliation{Harvard University, Cambridge, Massachusetts 02138} 
\affiliation{Division of High Energy Physics, Department of Physics, University of Helsinki and Helsinki Institute of Physics, FIN-00014, Helsinki, Finland} 
\affiliation{University of Illinois, Urbana, Illinois 61801} 
\affiliation{The Johns Hopkins University, Baltimore, Maryland 21218} 
\affiliation{Institut f\"{u}r Experimentelle Kernphysik, Universit\"{a}t Karlsruhe, 76128 Karlsruhe, Germany} 
\affiliation{Center for High Energy Physics: Kyungpook National University, Daegu 702-701, Korea; Seoul National University, Seoul 151-742, Korea; Sungkyunkwan University, Suwon 440-746, Korea; Korea Institute of Science and Technology Information, Daejeon, 305-806, Korea; Chonnam National University, Gwangju, 500-757, Korea} 
\affiliation{Ernest Orlando Lawrence Berkeley National Laboratory, Berkeley, California 94720} 
\affiliation{University of Liverpool, Liverpool L69 7ZE, United Kingdom} 
\affiliation{University College London, London WC1E 6BT, United Kingdom} 
\affiliation{Centro de Investigaciones Energeticas Medioambientales y Tecnologicas, E-28040 Madrid, Spain} 
\affiliation{Massachusetts Institute of Technology, Cambridge, Massachusetts  02139} 
\affiliation{Institute of Particle Physics: McGill University, Montr\'{e}al, Canada H3A~2T8; and University of Toronto, Toronto, Canada M5S~1A7} 
\affiliation{University of Michigan, Ann Arbor, Michigan 48109} 
\affiliation{Michigan State University, East Lansing, Michigan  48824}
\affiliation{Institution for Theoretical and Experimental Physics, ITEP, Moscow 117259, Russia} 
\affiliation{University of New Mexico, Albuquerque, New Mexico 87131} 
\affiliation{Northwestern University, Evanston, Illinois  60208} 
\affiliation{The Ohio State University, Columbus, Ohio  43210} 
\affiliation{Okayama University, Okayama 700-8530, Japan} 
\affiliation{Osaka City University, Osaka 588, Japan} 
\affiliation{University of Oxford, Oxford OX1 3RH, United Kingdom} 
\affiliation{Istituto Nazionale di Fisica Nucleare, Sezione di Padova-Trento, $^x$University of Padova, I-35131 Padova, Italy} 
\affiliation{LPNHE, Universite Pierre et Marie Curie/IN2P3-CNRS, UMR7585, Paris, F-75252 France} 
\affiliation{University of Pennsylvania, Philadelphia, Pennsylvania 19104}
\affiliation{Istituto Nazionale di Fisica Nucleare Pisa, $^y$University of Pisa, $^z$University of Siena and $^{aa}$Scuola Normale Superiore, I-56127 Pisa, Italy} 
\affiliation{University of Pittsburgh, Pittsburgh, Pennsylvania 15260} 
\affiliation{Purdue University, West Lafayette, Indiana 47907} 
\affiliation{University of Rochester, Rochester, New York 14627} 
\affiliation{The Rockefeller University, New York, New York 10021} 
\affiliation{Istituto Nazionale di Fisica Nucleare, Sezione di Roma 1, $^{bb}$Sapienza Universit\`{a} di Roma, I-00185 Roma, Italy} 

\affiliation{Rutgers University, Piscataway, New Jersey 08855} 
\affiliation{Texas A\&M University, College Station, Texas 77843} 
\affiliation{Istituto Nazionale di Fisica Nucleare Trieste/Udine, $^{cc}$University of Trieste/Udine, Italy} 
\affiliation{University of Tsukuba, Tsukuba, Ibaraki 305, Japan} 
\affiliation{Tufts University, Medford, Massachusetts 02155} 
\affiliation{Waseda University, Tokyo 169, Japan} 
\affiliation{Wayne State University, Detroit, Michigan  48201} 
\affiliation{University of Wisconsin, Madison, Wisconsin 53706} 
\affiliation{Yale University, New Haven, Connecticut 06520} 
\author{T.~Aaltonen}
\affiliation{Division of High Energy Physics, Department of Physics, University of Helsinki and Helsinki Institute of Physics, FIN-00014, Helsinki, Finland}
\author{J.~Adelman}
\affiliation{Enrico Fermi Institute, University of Chicago, Chicago, Illinois 60637}
\author{T.~Akimoto}
\affiliation{University of Tsukuba, Tsukuba, Ibaraki 305, Japan}
\author{M.G.~Albrow}
\affiliation{Fermi National Accelerator Laboratory, Batavia, Illinois 60510}
\author{B.~\'{A}lvarez~Gonz\'{a}lez}
\affiliation{Instituto de Fisica de Cantabria, CSIC-University of Cantabria, 39005 Santander, Spain}
\author{S.~Amerio$^x$}
\affiliation{Istituto Nazionale di Fisica Nucleare, Sezione di Padova-Trento, $^x$University of Padova, I-35131 Padova, Italy} 

\author{D.~Amidei}
\affiliation{University of Michigan, Ann Arbor, Michigan 48109}
\author{A.~Anastassov}
\affiliation{Northwestern University, Evanston, Illinois  60208}
\author{A.~Annovi}
\affiliation{Laboratori Nazionali di Frascati, Istituto Nazionale di Fisica Nucleare, I-00044 Frascati, Italy}
\author{J.~Antos}
\affiliation{Comenius University, 842 48 Bratislava, Slovakia; Institute of Experimental Physics, 040 01 Kosice, Slovakia}
% ...

\collaboration{CDF Collaboration\footnote{With visitors from $^a$University of Massachusetts Amherst, Amherst, Massachusetts 01003,
$^b$Universiteit Antwerpen, B-2610 Antwerp, Belgium, 
$^c$University of Bristol, Bristol BS8 1TL, United Kingdom,
$^d$Chinese Academy of Sciences, Beijing 100864, China, 
$^e$Istituto Nazionale di Fisica Nucleare, Sezione di Cagliari, 09042 Monserrato (Cagliari), Italy,
$^f$University of California Irvine, Irvine, CA  92697, 
$^g$University of California Santa Cruz, Santa Cruz, CA  95064, 
$^h$Cornell University, Ithaca, NY  14853, 
$^i$University of Cyprus, Nicosia CY-1678, Cyprus, 
$^j$University College Dublin, Dublin 4, Ireland,
$^k$Royal Society of Edinburgh/Scottish Executive Support Research Fellow,
$^l$University of Edinburgh, Edinburgh EH9 3JZ, United Kingdom, 
$^m$Universidad Iberoamericana, Mexico D.F., Mexico,
$^n$Queen Mary, University of London, London, E1 4NS, England,
$^o$University of Manchester, Manchester M13 9PL, England, 
$^p$Nagasaki Institute of Applied Science, Nagasaki, Japan, 
$^q$University of Notre Dame, Notre Dame, IN 46556,
$^r$University de Oviedo, E-33007 Oviedo, Spain, 
$^s$Simon Fraser University, Vancouver, British Columbia, Canada V6B 5K3,
$^t$Texas Tech University, Lubbock, TX  79409, 
$^u$IFIC(CSIC-Universitat de Valencia), 46071 Valencia, Spain,
$^v$University of Virginia, Charlottesville, VA  22904,
$^{dd}$On leave from J.~Stefan Institute, Ljubljana, Slovenia, 
}}
\noaffiliation

\preprint{CDF/PUB/ELECTROWEAK/PUBLIC/9368}
\preprint{Version 2.1}
\begin{abstract} 

We present a search for a Higgs boson decaying to two $W$ bosons 
in $\ppbar$ collisions at $\sqrt{s}=1.96$ TeV center-of-mass energy. The data
sample corresponds to an integrated luminosity of 3.0 fb${}^{-1}$ collected with the
CDF II detector. We find no evidence for production of a Higgs boson with mass between 110 and 200 $\GeV$,
and determine upper limits on the production cross section. For the mass of 160 $\GeV$, where the 
analysis is most sensitive,
the observed (expected) limit is 0.7 pb (0.9 pb) at 95\% Bayesian credibility level which is 1.7 (2.2) times 
the standard model cross section. 


\end{abstract}


\maketitle

\normalsize


The Higgs boson in the standard model (SM) breaks the electroweak $SU(2)_L \otimes U(1)_Y$
symmetry.  While this symmetry is now well established, 
the mechanism of the symmetry breaking has not yet been identified. Direct searches at the LEP experiments
have set a lower limit on the Higgs boson mass $\mh$  of 
114.4 $\GeV$ at 95\% C.L. in the context of the SM \cite{LEPHiggsDirectSearch}.
Precision measurements provide the indirect upper limit $\mh$ $<$ 144 $\GeV$ at 95\% C.L. through 
radiative corrections to the SM predictions of the particle masses and couplings  \cite{PrecisionEWK}. 
However, these indirect limits assume no significant contributions to the
radiative corrections due to as-yet unobserved processes.

In this letter, we report a search for the process $gg \rightarrow \HWW$
in a 3.0 fb${}^{-1}$ integrated luminosity sample of $\ppbar$ collisions at $\sqrt{s}=1.96$ TeV produced by the 
Fermilab Tevatron and collected by the CDF II detector. For an SM Higgs boson with a mass
that is not directly excluded by the LEP experiments, the dominant production 
mechanism at the Tevatron is gluon-gluon fusion which proceeds via a virtual 
top quark loop \cite{Djouadi:2005gi}.  For $\mh$ $>$ 135 $\GeV$, 
the SM Higgs boson decays primarily to the $WW^{*}$ \cite{Djouadi:2005gi}, 
where one of the final state
$W$ bosons is virtual for $\mh$ below two times the $W$ mass. 

The events are reconstructed in the $ll\nu\nu$ final state, whose branching fraction is 6.0\% 
of the $WW^{*}$ decays, where $l$ is either an electron $e$ or
a muon $\mu$, including those from $\tau$ leptons produced in the
$W$ decays. The SM Higgs boson branching fraction to $WW^{*}$  varies 
from 7.5\% at 115 $\GeV$ to 73.5\% at 200 $\GeV$ with a maximum of 96.5\% 
at $\approx$ 170 $\GeV$  \cite{Djouadi:2005gi}. 
Previous searches set limits ranging from 10 to 40 times
than the predicted SM rate, depending on the value of $\mh$   
\cite{Abazov:2005un}.


The CDF II detector is a multipurpose 
   solenoidal spectrometer surrounded by calorimeters and 
   muon detectors \cite{Acosta:2004yw}. The geometry 
is described using the azimuthal angle $\phi$ and the pseudorapidity $\eta
\equiv-\ln[\tan(\theta/2)]$, where $\theta$ is the polar angle with
respect to the proton beam axis (positive $z$-axis).  The transverse energy $\Et$ is $E\sin\theta$,
where $E$ is the energy associated with a 
calorimeter element or energy cluster. Similarly, $p_{T}$ is the track momentum component 
transverse to the beam line.


The events we consider must pass one of four online
selections, triggers, before being recorded. One electron trigger requires 
an electromagnetic (EM) energy
cluster in the central ($|\eta|<1.1$) calorimeter with $\Et >18$ GeV matched to 
a track found in the drift-chamber with $\pt > 8~\GeVc$.  A second electron trigger requires an EM energy
cluster with $\Et >20$ GeV in the forward ($1.2<|\eta|<2.0$) calorimeter and 
the missing transverse energy 
$\MET > 15$ GeV. The variable  
$\vec{\MET}$, used to infer the presence of neutrinos, is defined as  $ \sum_{i} E_{T,i} ~ \hat{n}_{T,i}
$ and $\MET\equiv|\vec{\MET}|$, where $\hat{n}_{T,i}$ is the transverse component of the unit
vector pointing from the interaction point to calorimeter element
$i$. Muon triggers are based on track segments
in the muon chambers matched to a drift-chamber track with $\pt > 18~\GeVc$. Trigger efficiencies are
measured using leptonic $W$ and $Z$ data samples  \cite{Acosta:2004uq}.


To improve the signal acceptance while maintaining acceptable 
background rejection for the $W$+jets and $W\gamma$ processes where a jet or $\gamma$ is 
misidentified as a lepton, we use a modified version of the 
lepton identification strategy developed for the {\it WZ} observation analysis
 \cite{Abulencia:2007tu}. Candidate leptons are separated into six
mutually exclusive categories: two for electrons; three for muons; and
one for tracks that extrapolate outward to detector regions with insufficient calorimeter
coverage for energy measurement. The electron categories
are central ($|\eta|<1.1$) using a drift-chamber-based tracking algorithm and
forward ($1.2<|\eta|<2.0$) using a silicon-detector-based tracking algorithm.
One of the muon categories uses the muon chambers 
and other two use tracks matched with energy deposits 
consistent with minimum ionization in the central or forward calorimeters.

All lepton candidates are required to be isolated such that the sum
of the $E_{T}$ for the calorimeter elements in a cone of $\Delta R =
\sqrt{(\Delta\eta)^2 + (\Delta\phi)^2} < 0.4$ around the lepton is
less than 10\% of the $\Et$ (for electrons) or $\pt$ (for muons and track
lepton candidates). For lepton types in the central region
where the track finding efficiency is sufficient, we also apply a
track-based isolation criterion which requires there is no more than 10\% of the
electron $\Et$ or muon $\pt$ in other tracks within a cone of $\Delta R <
0.4$ around the lepton track.


The Higgs boson candidates are selected from events with
exactly two  lepton candidates.
At least one lepton is 
required to match a trigger lepton candidate and have $\Et > 20$ GeV ($\pt >
20~\GeVc$) for electrons (muons). We loosen this requirement to $10$
GeV ($\GeVc$) for the other lepton to increase the kinematic
acceptance, particularly for lower $\mh$ where one $W$ is off-shell below 
the $W$ boson mass and produces a lower $\pt$ lepton. 
We also require a dilepton invariant mass $m_{\ell\ell} >
16~\GeV$ to suppress misidentified multijet events.

Aside from $\HWW$ production, other SM processes that can 
lead to two high-$\pt$ leptons include Drell-Yan (DY),
$\ttbar$, $\WW$, $\WZ$, and $\ZZ$ production, and $\W$ production
in association with a photon ($W\gamma$) or a jet ($W$+jets) misidentified as a lepton.
The $\ttbar$ contribution is
suppressed by requiring fewer than two reconstructed jets with $\Et >
15$ GeV and $|\eta| < 2.5$ in the event. This allows for up to one jet due to inital 
state radiation. The DY background is
suppressed by requiring sufficiently large $\MET$ in the event to
remove contributions from mismeasured leptons or jets. This is
achieved by requiring $\METrel > 25~\mathrm{GeV}$, where 
\begin{equation*}
  \METrel \equiv \left\{ 
  \begin{array}{ll} 
    \MET                       & \mbox{if } \MetDeltaPhi > \frac{\pi}{2} \\
    \MET\sin{\MetDeltaPhi}   & \mbox{if } \MetDeltaPhi < \frac{\pi}{2} \\
  \end{array} \right\},
\end{equation*}
and $\MetDeltaPhi$ is the angle between the $\vecMET$ direction
and the nearest lepton or jet.  The observed $\MET$ is corrected for 
muons and track-only lepton candidates, because they do not deposit all 
of their energy in the calorimeter. 

The acceptances for the $\HWW$, $\WW$, $\WZ$, $\ZZ$, $W\gamma$, DY, and $\ttbar$
processes are determined using simulated data. Events are simulated with the
{\sc mc@nlo} program for $\WW$  \cite{Frixione:2002ik}, {\sc pythia}
for $\HWW$, DY, $\WZ$, $\ZZ$, and $\ttbar$  \cite{Sjostrand:2006za}, and the
generator described in Ref. \cite{Baur:1992cd} for $W\gamma$.
The response of the CDF II detector is then estimated with 
a {\sc geant-4}-based simulation  \cite{GEANT} to which 
an efficiency correction of up to 10\% per lepton is applied
based on measurements of the lepton reconstruction and identification 
efficiencies using observed $\Z\rightarrow\ell^{+}\ell^{-}$ events.
An additional
correction is applied to the $W\gamma$ background estimate based on a
measurement in data of the efficiency of requirement that supresses the fraction of  
photon conversions identified as electrons. The
$W$+jets contribution is estimated from data by extrapolating from
a sample of events that contain an identified lepton and a jet
$j_l$ containing a track or EM energy cluster similar to those
required in the lepton identification. The contribution of each event
to the total yield is scaled by the probability that the $j_l$ is
identified as a lepton. This probability $p(p_T;j_l)$ is determined for
each lepton type as a function of the $j_l$ $p_T$ using multijet events 
collected with jet-based triggers. A correction to $p(p_T;j_l)$ is
applied for the small real lepton contribution using Monte Carlo
simulation of single $\W$ and $\Z$ boson production. 

Based on the procedure above, we expect $768 \pm 91$ background 
events and observe 779 in the selected region. The expected composition of the 
background is  
      356  $\pm$   49 {\it WW},
      24.9 $\pm$ 3.9  {\it WZ},
      21.8 $\pm$ 3.5  {\it ZZ},
      25.5 $\pm$ 5.0  $t\bar{t}$,
      138  $\pm$ 31   DY,
      90.5 $\pm$ 24.1 $W\gamma$, and 
      111  $\pm$ 27   $W$+jets, where the indicated uncertainties include the
systematic uncertainties described below.
As a cross-check of the background model, we measure the fraction of $\qqbar\rightarrow WW$ 
events in the sample with a similar method to the signal extraction described below
and find it to be consistent with the expectation.
The largest predicted SM
Higgs boson yield is 11.6 events at $\mh$=160 $\GeV$ which results from an acceptance of $16\%$. 
Table \ref{tab:yieldvsmh} shows the 
dependence of expected yield on $\mh$. 
Since the level of non-$WW^{(*)}$ background depends on the lepton identification
categories, this information is used to divide the sample into  
high and low signal-to-background ($S/B$) classes.

\begin{table}
\caption{Expected Higgs boson yield as a function of $\mh$.}
\begin{tabular}{lcccccccccc}
\hline
\hline
$\mh$ ($\GeV$) & 110 & 120 & 130 & 140 & 150 & 160 & 170 & 180 & 190 & 200 \\
Expected yield &  0.5      &      1.9      &      4.3      &      7.0      &      9.3      &      11.6      &      11.0      &      9.0      &      6.4      &      5.1 \\
\hline
\hline
\end{tabular}
\label{tab:yieldvsmh}
\end{table}


 After selection, the dominant background 
is the $\qqbar\rightarrow WW$ process which differs from the signal
process only by spin, production mechanism, and resonant structure. 
Because of the two neutrinos in the final state, a simple mass peak cannot
be used to isolate the resonant $gg\rightarrow \HWW$ process from the 
backgrounds. Instead, we combine
two different multivariate techniques to differentiate signal and background.
One is a matrix element (ME) technique, which uses an event-by-event calculation
of the probability density for each contributing process to produce the 
observed event. If all details of the collision properties and the detector
response are modeled in the ME calculation, this method provides the optimal sensitivity
to the signal. However, there are several approximations used in the calculations:
theoretical differential cross sections are only implemented to leading order,
a simple parameterization of the detector response is used, and for some small
({\it WZ} and $\ttbar$) or difficult to model (DY) backgrounds, we do not calculate
a probability density.
In order to improve these approximations, we extend the ME calculation with 
a neural network (NN) which exploits the more complete model 
implemented in the simulated data and $W$+jets model. 

\begin{figure}
\includegraphics[width=0.4\textwidth]{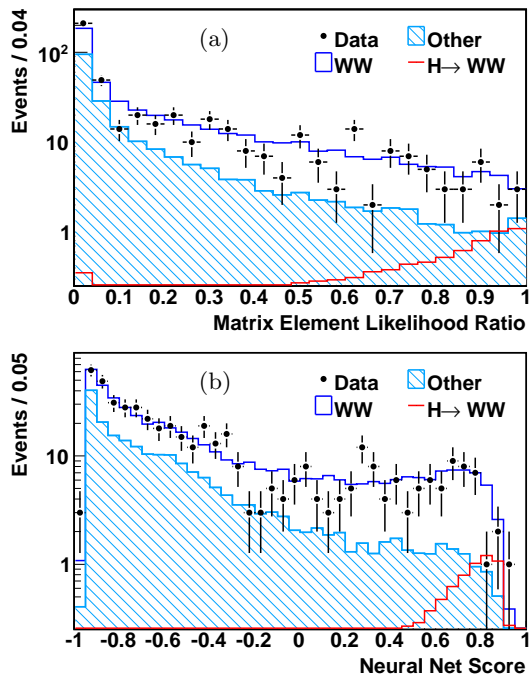}
\unitlength=0.96\linewidth
\put(-0.55,1.02){(a)}
\put(-0.55,0.47){(b)} 

\caption{The likelihood ratio distribution from a) the ME discriminate alone and b) the 
         full NN score for $\mh=160$ $\GeV$. The Higgs boson distribution is normalized 
         to the SM expectation. The distributions are shown only for the high $S/B$
         class, which provides the majority of the sensitivity. }

\label{fig:discriminants}

\end{figure}

The event probability density for the ME method is
\begin{equation*}
\label{eqn:master_prob}
P(\vec{x}_{obs}) =
  {1 \over \langle\sigma\rangle }
  \int \frac {d \sigma_{LO} (\vec{y}) }{ d\vec{y} }
  \epsilon (\vec{y}) G(\vec{x}_{obs},\vec{y}) d\vec{y},
\end{equation*}
where the elements of $\vec{y}$ ($\vec{x}_{obs}$) are the true (observed) values
of the lepton momenta and $\MET$, $d \sigma_{LO} / d\vec{y}$ is the
parton level differential cross section \cite{Campbell:1999ah},
$\epsilon(\vec{y})$ is a parameterization of detector acceptance and efficiency function, and
$G(\vec{x}_{obs},\vec{y})$ is the transfer function representing the detector
resolution and a {\sc pythia}-based estimate of transverse momentum of the $\ell\ell\MET$
system due to the initial state radiation. The constant $\langle\sigma\rangle$ normalizes the total
event probability to unity. This calculation integrates the theoretical 
differential cross section over the 
missing information due to two unobserved neutrinos 
in the final state. We form a likelihood ratio discriminant which
is the signal probability divided by the sum of signal and background 
probabilities $LR_{\HWW}(\vec{x}_{obs})\equiv\frac{P_{H}(\vec{x}_{obs})}{P_{H}(\vec{x}_{obs})+ \sum_i k_i P_{i}(\vec{x}_{obs})}$
where $k_i$ are the expected background fractions of 
{\it WW}, {\it ZZ}, $W\gamma$, and $W$+jets.
The $LR$ distributions are shown in Figure
\ref{fig:discriminants}(a). 
Additional ME likelihood ratios $LR_{WW}$, $LR_{ZZ}$, $LR_{W\gamma}$, 
and $LR_{W+\textrm{jets}}$ are defined analogously to $LR_{\HWW}$.

For the final results, an NN discriminant
is used to extend the ME calculation using as input the ME likelihood ratios in addition 
to various kinematic variables.
For each of the Higgs boson masses investigated, an NN is trained 
on signal events and an appropriately weighted composition of background 
events.
The NN classifies events as signal or background based on the inputs
$LR_{\HWW}$, $LR_{WW}$, $LR_{ZZ}$, $LR_{W\gamma}$, $LR_{W+\textrm{jets}}$,
$\Delta \phi_{\ell\ell}$, $\Delta R_{\ell\ell}$, $m_{\ell\ell}$, $\MET$, 
$\MetDeltaPhi$, and 
$\METrel$, where $\Delta \phi_{\ell\ell}$ and $\Delta R_{\ell\ell}$ are the 
separation between the two leptons in $\phi$ and $\Delta R$,
respectively. 
We find that the most discriminating input variables are $LR_{\HWW}$, $\Delta R_{\ell\ell}$, and $\METrel$.
An example of the NN output is shown 
in Figure~\ref{fig:discriminants}(b).
The presented results use the {\sc neurobayes} \cite{Feindt:2006pm} program.
Comparable results are also obtained
using the {\sc tmva-mlp} \cite{Hocker:2007ht} program, demonstrating the robustness
of the technique.

The systematic uncertainties associated with the  Monte Carlo simulation 
affect all of the signal and background expectations similarly, except for $W$+jets. The
uncertainties from the lepton selection and trigger efficiency
measurements are propagated through the analysis, giving uncertainties
from $1.4\%$ to $2.0\%$ and $2.1\%$ to $7.1\%$ for the respective
efficiencies of the different signal mass points and background processes. The
detector acceptance variation due to parton-distribution fuction (PDF) uncertainties is assessed to
be $1.9\% - 4.1\%$ using the 20 pairs of PDF sets described in
Ref. \cite{Kretzer:2003it}.  We assign a 6\% luminosity uncertainty to
the signal and background estimates that are obtained from simulation
 \cite{Acosta:2002hx}. The cross section uncertainties are 10\% for
{\it WW}  \cite{Campbell:1999ah}, {\it WZ}  \cite{Campbell:1999ah}, {\it ZZ}  \cite{Campbell:1999ah}, and
$W\gamma$  \cite{Baur:1997kz}, and 15\% for $\ttbar$
 \cite{Kidonakis:2003qe}. Based on a comparison of simulated 
{\it WW} events generated with {\textsc mc@nlo} and {\sc pythia}, we 
assign a systematic uncertainty on the acceptance due to higher order QCD effects
of 5.5\% for {\it WW} events and 10\% for the other modes which are only simulated at leading order.

The systematic uncertainty on the $W$+jets background is estimated to be 
24\% from differences in the observed
probability that a jet is identified as a lepton for jets collected
using different jet $\Et$ trigger thresholds. These variations
correspond to changing the parton composition of the jets and the
relative amount of contamination from real leptons. 
Because only the $\METrel$ requirement suppresses the DY background,
there is an uncertainty due to the $\MET$ resolution modeling, which 
is estimated to be 20\% based on comparisons of the data and Monte Carlo simulation in a sample of
dilepton events. For the $W\gamma$ background contribution, there is
an additional uncertainty of 20\% from the detector material
description and photon-conversion veto efficiency. 

\begin{figure}
\includegraphics[width=0.4\textwidth]{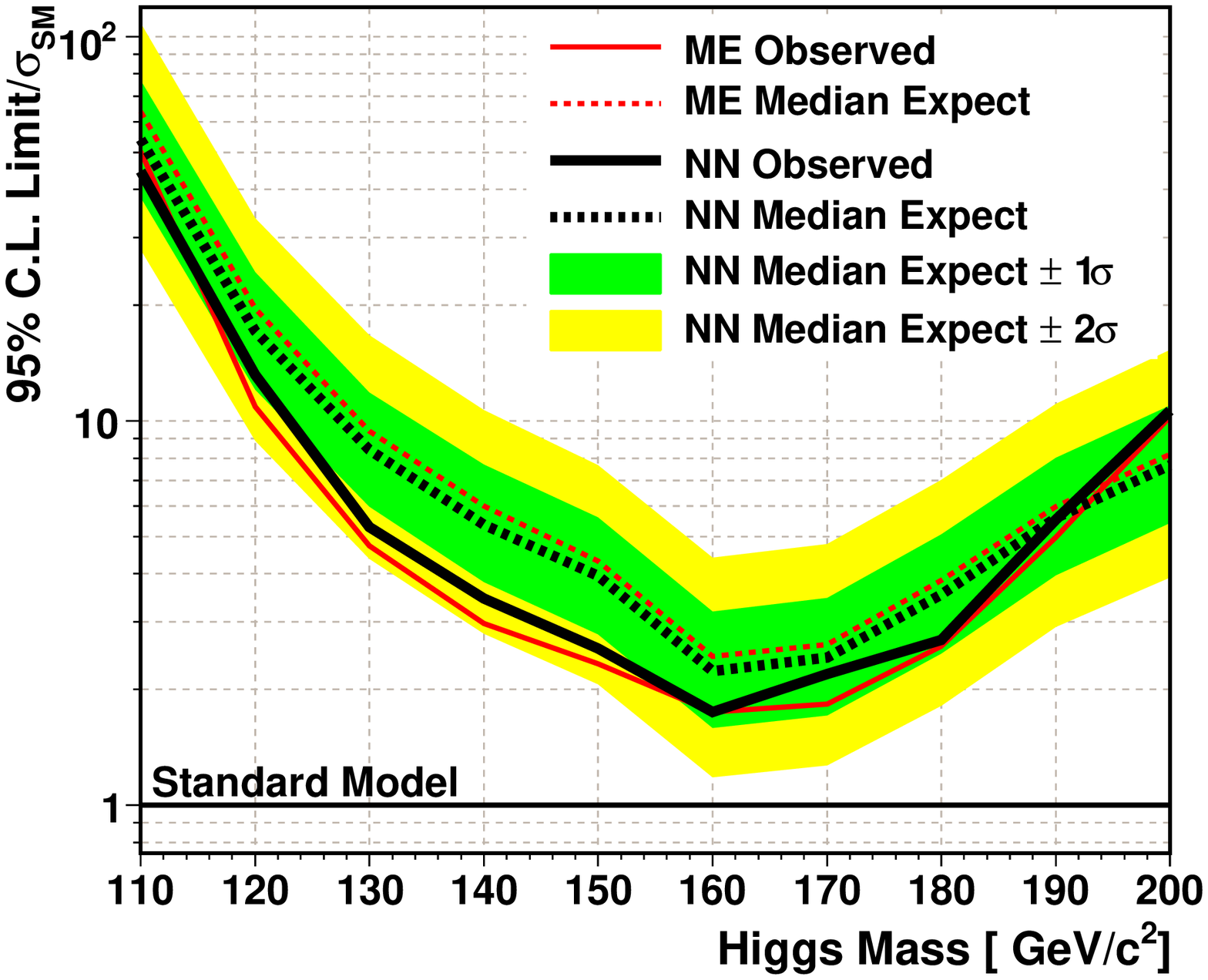}
\caption{Upper limits on $\sigma(gg\rightarrow H)\times{\cal{B}}(\HWW)/\sigma_{\rm SM}(gg\rightarrow H)\times{\cal{B}_{\rm SM}}(\HWW)$ as a function of $\mh$.}
\label{fig:limits}
\end{figure}

A Bayesian credibility level (C.L.) is calculated for each $\mh$ hypothesis
based on the combined binned likelihood of the discriminant distributions for
the high and low $S/B$ samples.
A posterior density is obtained by multiplying this likelihood 
by Gaussian prior densities for the background normalizations and systematic 
uncertainties leaving $\sigma(gg\rightarrow H)\times{\cal{B}}(\HWW)$ 
with a uniform prior density.
A 95\% C.L. limit is then determined such that 95\% of the posterior density for
$\sigma(gg\rightarrow H)\times{\cal{B}}(\HWW)$ falls below
the limit. Limits as a fraction of the SM cross section 
$\sigma_{\rm SM}(gg\rightarrow H)$ are calculated by including $\sigma_{\rm SM}(gg\rightarrow H)$
as a parameter whose prior density is determined by the 10\% theoretical uncertainty 
of its next-to-next-to-leading-log prediction \cite{Catani:2003zt}. The resulting cross section limits for both
discriminants are shown in Table \ref{tab:limits} and Figure \ref{fig:limits}. 

\begin{table}
\caption{Expected and observed upper limits on $\sigma(gg\rightarrow H)\times{\cal{B}}(\HWW)$
and $\sigma(gg\rightarrow H)\times{\cal{B}}(\HWW)/\sigma_{\rm SM}(gg\rightarrow H)\times{\cal{B}_{SM}}(\HWW)$ as a function of $\mh$.}
\begin{tabular}{lcccccccccc}
\hline
\hline
$\mh$ ($\GeV$) & 110 & 120 & 130 & 140 & 150 & 160 & 170 & 180 & 190 & 200 \\
\hline
\multicolumn{11}{c}{Using Matrix Element Only} \\
Expected (pb)     &     3.6 &   2.6 &   2.2 &   1.9 &   1.5 &   0.9 &   0.9 &   1.1 &   1.2 &   1.3 \\
Observed (pb)     &     2.8 &   1.5 &   1.1 &   0.9 &   0.8 &   0.7 &   0.6 &   0.7 &   1.0 &   1.5 \\
Expected/SM       &    63.7 &  19.6 &   9.4 &   6.0 &   4.3 &   2.4 &   2.6 &   3.8 &   6.0 &   8.2 \\
Observed/SM       &    50.3 &  10.9 &   4.7 &   3.0 &   2.3 &   1.7 &   1.8 &   2.6 &   5.0 &  10.3 \\
\hline
\multicolumn{11}{c}{Using Neural Net Discriminator} \\
Expected (pb)     &       3.0 &   2.3 &   1.9 &   1.7 &   1.4 &   0.9 &   0.8 &   1.0 &   1.1 &   1.2 \\
Observed (pb)     &       2.5 &   1.7 &   1.2 &   1.1 &   0.9 &   0.7 &   0.7 &   0.7 &   1.0 &   1.6 \\
Expected/SM       &      54.0 &  17.1 &   8.4 &   5.4 &   3.9 &   2.2 &   2.4 &   3.5 &   5.6 &   7.7 \\
Observed/SM       &      44.6 &  13.2 &   5.3 &   3.5 &   2.6 &   1.7 &   2.2 &   2.7 &   5.5 &  10.6 \\
\hline
\hline
\end{tabular} 
\label{tab:limits}
\end{table}

In conclusion, we have presented limits on the production of a Higgs boson
through gluon fusion followed by its decay to a pair of $W$ bosons.
A combination of matrix element and neural network techniques is used
to discriminate signal from background. Studies using the two 
techniques independently achieve consistent results with a sensitivity
approximately 0.1$\sigma_{\rm SM}$ worse than the combination at $\mh=160~\GeV$.
The consistency of results obtained with different algorithms
provides evidence of the robustness of the multivariate techniques.
At the most sensitive value of $\mh=160~\GeV$,
the observed limit is 1.7 times the SM prediction where the median expected limit
is 2.2, corresponding to a downward fluctuation slightly larger than one standard deviation.
Compared to an optimized selection and a likelihood based on the $\Delta \phi_{\ell\ell}$ variable, 
the multivariate discriminators gain a factor of 1.7 to 2.5 in effective integrated luminosity depending on $m_H$.
This measurement also constrains alternative models in which the $gg\rightarrow H$ 
coupling is enhanced by additional particles in the virtual loops of the production amplitude  
\cite{Manohar:2006gz}.


We thank the Fermilab staff and the technical staffs of the participating institutions for their vital contributions. This work was supported by the U.S. Department of Energy and National Science Foundation; the Italian Istituto Nazionale di Fisica Nucleare; the Ministry of Education, Culture, Sports, Science and Technology of Japan; the Natural Sciences and Engineering Research Council of Canada; the National Science Council of the Republic of China; the Swiss National Science Foundation; the A.P. Sloan Foundation; the Bundesministerium f\"ur Bildung und Forschung, Germany; the Korean Science and Engineering Foundation and the Korean Research Foundation; the Science and Technology Facilities Council and the Royal Society, UK; the Institut National de Physique Nucleaire et Physique des Particules/CNRS; the Russian Foundation for Basic Research; the Ministerio de Ciencia e Innovaci\'{o}n, Spain; the Slovak R\&D Agency; and the Academy of Finland. 


\end{document}